# Nanoscale Tantalum Layer Controlling the Magnetic Coupling between Two Ferromagnetic Electrodes via Insulator of a Magnetic Tunnel Junction

Pawan Tyagi[1,2]* and Tobias Goulet[1]

[1]Mechanical Engineering, University of the District of Columbia, Washington DC-20008. USA

[2]Chemical and Materials Engineering, University of Kentucky, Lexington, KY-40566, USA

*Corresponding Author: ptyagi@udc.edu

**ABSTRACT:** Ability to tailor the nature of the magnetic coupling between two ferromagnetic electrodes can enable the realization of new spintronics device systems. This paper discusses our finding that deposition of an ultrathin tantalum (Ta) on the NiFe top electrode reversed the nature of inter-ferromagnetic electrode coupling. We observed that the deposition of ~ 5 nm Ta on the top of a magnetic tunnel junction with Ta( 2 nm)/Co(5 nm)/NiFe (5 nm)/AlOx( 2 nm)/NiFe (10-15 nm) configuration changed the magnetic coupling between two ferromagnetic electrodes from antiferromagnetic to ferromagnetic. We investigated Ta effect using multiple magnetic characterizations like ferromagnetic resonance, magnetometry, and polarized neutron reflectometry. Ferromagnetic resonance characterization was very sensitive for detecting the changes in magnetic coupling via the insulating spacer. This simple approach of adding Ta film to alter the magnetic coupling can impact the other burgeoning areas like molecular spintronics. We found that preexisting magnetic coupling between two ferromagnetic electrodes impacted the resultant magnetic properties of magnetic tunnel junctions based molecular spintronics devices.

**I INTRODUCTION:**

Tailoring the nature of the magnetic coupling between two ferromagnetic electrodes has been the topic of intense interest [1, 2]. Ability to change the inter-ferromagnetic electrode coupling can lead the development of new device forms and materials [2]. For instance, nanoscale spintronics devices focus on maneuvering the nature and strength of the inter-ferromagnetic electrode coupling (IFMEC)[2]. To date, three key approaches have been employed to tailor the IFMEC. The first approach involves inserting the nanoparticles between two ferromagnetic electrodes [1]. The second method involves changing the thickness of nonmagnetic spacers between two ferromagnetic electrodes [3]. The third method requires changing of the nonmagnetic spacer material between two ferromagnetic electrodes. However, these approaches are very challenging to implement. For instance, controlling the distribution and sizes of nanoclusters between two ferromagnetic electrodes is very challenging to exercise and difficult to reproduce [1]. Similarly, tailoring the spacer thickness to sub-nm scale and even changing the spacer material altogether requires intensive device optimization[2, 4]. An approach that does not physically affect the spacer between two ferromagnetic electrodes and easy to implement can lead to new opportunities. Recently, magnetic tunnel junction based molecular spintronics devices (MTJMSDs) were developed [5-7]. One can alter the IFMEC by adding Ta top layer before transforming a magnetic tunnel junction into a MTJMSD. In this paper, we first discuss the role of Ta on IFMEC. We also discussed the impact of preexisting IFMEC on the magnetic properties of the MTJMSD.

**II EXPERIMENTAL DETAILS:**

To investigate Ta effect on IFMEC we employed various physical property measurement techniques such as Ferromagnetic resonance (FMR), magnetometry, and polarized neutron reflectivity (PNR). For the PNR measurements, unpatterned MTJ and MTJ-Ta samples were employed. The reason for utilizing unpatterned samples was based on the strong effect from the uncovered substrate that strongly impacted the measurement and modeling accuracy. The FMR and magnetometry study utilized patterned tunnel junctions. The FMR and magnetometry methods are sensitive towards the magnetic property of the materials. Unlike PNR the FMR and magnetometry do not get influenced by the nonmagnetic substrate. Also, utilizing patterned tunnel junction was necessary for making molecular devices for the FMR and magnetometry study. It is notworthy that edge effects become prominent for submicron or nm scale magnetic features [8]. To avoid the undesirable impact of edges we produced the MTJ and MT-Ta with several tens of micron area. We fabricated an array of ~7000 patterned magnetic tunnel junctions per sample. Every magnetic tunnel junction was ~5 μm in diameter and ~10 μm distance from the neighboring magnetic tunnel junctions. We utilized an oxidized silicon substrate and performed photolithography to produce a photoresist layer with an array of micro-cavities. For this study, Shipley® (S1813) photoresist was spin coated on an oxidized silicon wafer piece at 3000 rpm speed. This spin coated photoresist film was baked at 90 $^0$C for one minute. Subsequently, the photoresist was exposed to UV light through a photomask containing an array pattern. We developed the photoresist film in MF 319 Microposit developer to produce an array of microcavities in the photoresist film. These microcavities were filled with multiple thin films to create an array of magnetic tunnel junctions. Tantalum (Ta), cobalt (Co), alumina(AlOx), and the alloy of NiFe alloy with 80% nickel were sputter deposited with AJA International sputtering machine. We first sputter deposited Ta(2 nm)/Co(5 nm)/NiFe (5 nm)/AlOx( 2 nm)/NiFe (10-15 nm) thin film configuration. This magnetic tunnel junction configuration is named as MTJ. To prepare the sample for the study of Ta impact on IFMEC MTJ-Ta samples were produced. To create MTJ-Ta samples, we additionally sputter deposited 5 nm Ta on the top of MTJ configuration. A resultant MTJ-Ta sample had Ta(2 nm)/Co(5 nm)/NiFe (5 nm)/AlOx( 2nm)/NiFe (10-15 nm)/Ta (5 nm) configuration. The ~ 5 nm Ta top layer thickness ensured a conformal film on the top NiFe electrode. We noted that deposition of NiFe on the alumina (AlOx) generally produced as high as 3-5 nm RMS roughness. The high roughness is inevitable due to the amorphous nature of AlOx [8]. With such high roughness Ta < 3 nm was unable to form conformal films. Hence, Ta with thickness around 5 nm was a good choice. On the other hand increasing Ta thickness also reduced device yield, presumably due to higher mechanical stresses. Mechanical stresses have been found to create tunnel barrier failures [9]. These MTJ and MTJ-Ta samples were characterized by the X-band Bruker EMX300 FMR and NanoOsc Phase FMR over 2-17 G Hz frequency range at room temperature. Magnetization studies were performed with Quantum Design PPMS SQUID magnetometer at 150 K. Low temperature was chosen to avoid noise in the magnetization study. PNR study was performed at 150 K and at 150 mT magnetic field at National Institute of Standards and Technology, Gaithersburg USA. The light reflectivity studies were performed with Semiconsoft ® Mprobe thin film measurement system at room temperature.

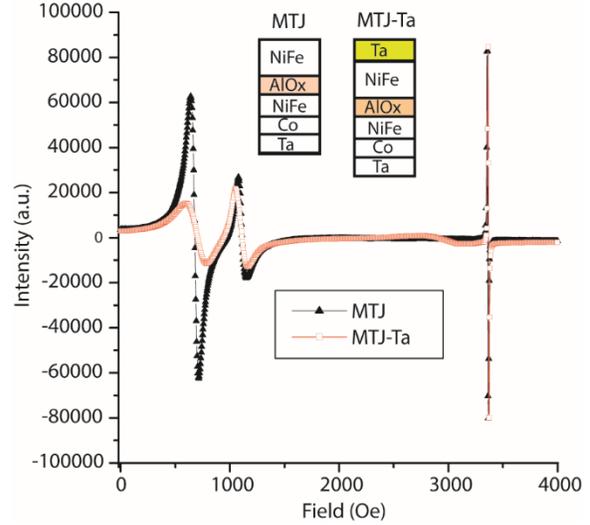

Fig. 1. FMR spectra recorded on MTJ and MTJ-Ta samples.

**III RESULTS AND DISCUSSIONS**:

We first studied the IFMEC on MTJ and MTJ-Ta. It is noteworthy that two magnetic structures separated by nm gap exhibit ferromagnetic or antiferromagnetic couplings [10]. FMR is a powerful tool to study the characteristics of magnetic coupling between two magnetic structures, especially two ferromagnetic films [11]. Our FMR studies revealed a striking difference between MTJ and MTJ-Ta (Fig. 1). Under the identical experimental conditions, both samples showed two distinct FMR modes. For both samples, an in-plane DC magnetic field up to 4000 Oe and 9.75G Hz microwave was applied to study the resonance modes. Before every measurement, the cavity's spectra was checked for the background signal at fivefold higher gain than that used for the MTJ and MTJ-Ta samples. We also investigated if the manual error in the sample alignment with respect to the direction of the magnetic field could impact the intensity of FMR modes. The FMR spectra of a sample did not change noticeably within the ±10º variation on the magnetic field direction. We also utilized the FMR response from the graphite tape as a control sample to ensure that experimental conditions were identical for the MTJ and MTJ-Ta (Fig. 1). During the FMR study of MTJ and MTJ-Ta, the graphite tape produced a delta function type resonance peak at 3367±4 Oe (Fig. 1). In our study, we utilized the same graphite tape to mount the MTJ and MTJ-Ta samples for the FMR study. This sharp resonance peak from the graphite tape was statistically identical for the MTJ and MTJ-Ta samples. Invariance of the graphite tape's signal suggests that experimental conditions for measuring MTJ and MTJ-Ta were identical. The reproducibility of the graphite tape's resonance signal ensured the robustness and reproducibility of microwave power, DC magnetic field, and losses due to the measurement system, etc.

The MTJ sample showed acoustic mode (higher intensity resonance peak) before the optical mode (lower intensity peak). According to FMR theory [12],[13] two ferromagnetic electrodes of the MTJ sample are antiferromagnetically coupled. On the other hand, the two resonance peaks from MTJ-Ta sample were significantly different as compared to the FMR peaks from MTJ. It appears that presence of Ta on the top of NiFe ferromagnetic electrode reduced the intensity of the first resonance peak. As a result, MTJ-Ta exhibited smaller intensity resonance mode (optical mode) appearing before the higher intensity resonance (acoustic mode). This particular form of the FMR spectra from MTJ-Ta is indicative of ferromagnetic coupling between the two ferromagnetic electrodes[12],[13].

We estimated the strength of exchange coupling between two ferromagnetic electrodes of the MTJ and MTJ-Ta to be of the similar magnitude. We estimated the order of magnitude of the magnetic coupling by two ways. (a) First, we evaluated the slope of the lines joining the two resonance modes of the MTJ and MTJ-Ta. It is noteworthy that zero slop means no interaction between the two ferromagnetic electrodes. The FMR modes recorded on the isolated top and bottom ferromagnetic electrodes are uncoupled (Fig. 2). However, for MTJ and MTJ-Ta the slope of the line between two modes is roughly the same. (b) The estimation of magnetic coupling strength is also possible from the difference in tunnel junction's mode positions with respect to the mode positions from the isolated ferromagnetic electrodes. It is noteworthy that tunnel junction's mode positions shift as a function of the magnetic coupling strength between two electrodes [10, 14]. We noticed that resonance position for the MTJ and MTJ-Ta was only ~45 Oe less as compared to the resonance magnetic field of the Ta/Co/NiFe bottom electrode grown in isolation. We surmise that addition of Ta only affected the nature of IFMEC, not its magnitude. It is also noteworthy that adding Ta appears to reduce the intensity of acoustic mode, which appeared close to the top NiFe electrode's resonance position (Fig. 2).

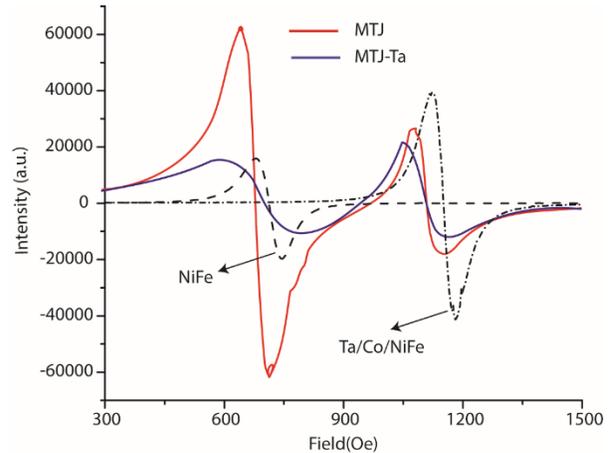

Fig. 2. FMR spectra of MTJ and MTJ-Ta samples with respect to FMR peaks from the isolated Ta/Co/NiFe and NiFe electrodes.

We also investigated the difference in magnetization data obtained from the MTJ and MTJ-Ta. We found that both samples produced almost identical magnetization loops (Fig. 3). However, MTJ-Ta was relatively less sloped in the unsaturation state (Inset of Fig. 3). This Ta induced subtle difference in the magnetization loop affirms two important points: (i) For MTJ the antiferromagnetic

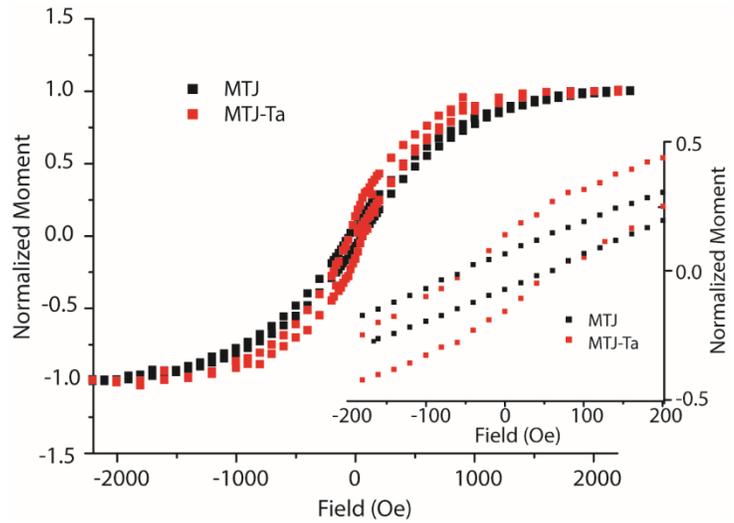

Fig.3: .Magnetic field vs. normalized moment for ±2000 Oe range. Inset image show zoomed in image of magnetization curve for ± 200 Oe.

IFMEC is not strong otherwise there could be a significant changes in the magnetization loop [15]; (ii) magnetization data for MTJ-Ta showed moderate increase in the magnetic moment between saturation states as compared to that of MTJ (Fig. 3a). We surmise that this moderate increase in magnetization is due to the emergence of ferromagnetic coupling. If the addition of Ta did not affect the coupling, then magnetization loop should have been the same for the MTJ and MTJ-Ta samples. On the other hand, if the addition of Ta enhanced the antiferromagnetic coupling, then the magnetic moment in the unsaturation region should have been reduced. Hence, the current form of the magnetization data asserts with the FMR data (Fig. 1), which indicates that the addition of Ta produced ferromagnetic coupling between two ferromagnetic electrodes.

To investigate the mechanism behind the Ta effect, we conducted FMR study on tunnel junctions where the ferromagnetic electrodes on both sides of the AlOx tunneling barrier were the same. We studied the FMR response from a tunnel junction comprising of Ta(2 nm)/Co(5 nm)/NiFe (5 nm)/AlOx( 2 nm)/NiFe (10 nm)  (Fig. 4). This configuration essentially showed that bottom ferromagnetic electrode of the MTJ and MTJ-Ta is present on the both sides of AlOx insulator. The FMR spectra for this tunnel junction configuration revealed antiferromagnetic coupling between the two ferromagnetic electrodes via AlOx insulator. The existence of this antiferromagnetic coupling is evident from the locations of acoustic and optical modes [13]. In this case, acoustic mode appeared at 1025 Oe, and optical mode existed at 3125 Oe (Fig. 4). The resonance peaks for this configuration are ~2000 Oe apart, whereas two resonance modes for the MTJ and MTJ-Ta were positioned at a gap of ~340 Oe only.

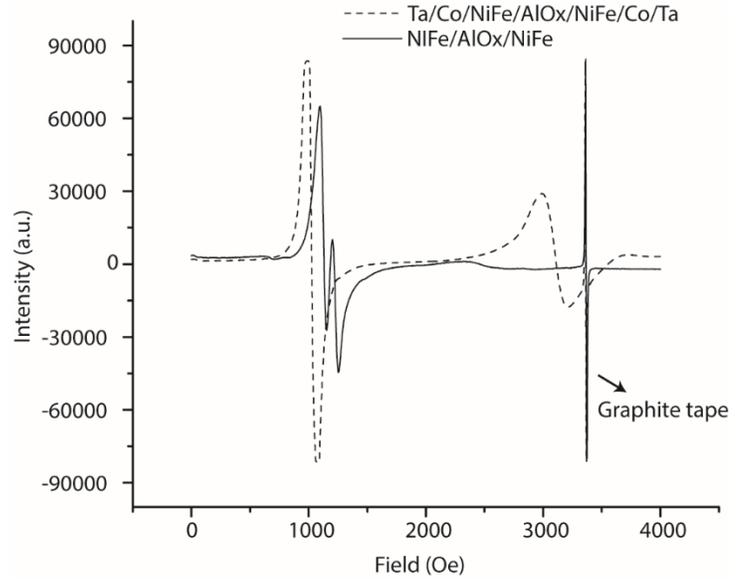

Fig. 4: FMR of magnetic tunnel junctions with Ta(2 nm)/Co(5 nm)/NiFe (5 nm)/AlOx( 2 nm)/NiFe (10 nm) and NiFe(10 nm)/AlOx (2 nm)/NiFe (10 nm) configurations.

We also studied FMR on the tunnel junction with NiFe (10 nm)/AlOx (2 nm)/NiFe (10 nm) configuration (Fig. 4). It is noteworthy that the top and bottom ferromagnetic electrodes of this tunnel junction are similar to the ferromagnetic film utilized in the MTJ and MTJ-Ta. This tunnel junction also exhibited antiferromagnetic coupling between two NiFe ferromagnetic electrodes. For this configuration, acoustic mode appeared before the optical mode. Acoustic mode appeared at ~1130 Oe, and optical mode appeared at 1230 Oe. Hence, the difference between two modes for this tunnel junction was only ~100 Oe that is significantly smaller than that for MTJ and MTJ-Ta, i.e. 340 Oe. This tunnel junction with similar ferromagnetic electrodes also possessed antiferromagnetic coupling [13]. Adding Ta on the top of this tunnel junction did not produce the change in the IFMEC. It appears that impact of Ta is pronounced for the magnetic tunnel junction with the dissimilar magnetic electrodes.

To understand the impact of Ta along the depth of the MTJ, we conducted PNR studies. We hypothesized that magnetic attributes of the MTJ's top electrode should be affected by the Ta layer. We employed Polarized Neutron Reflectometry (PNR) to investigate the difference in magnetic attributes of the MTJ and MTJ-Ta. Under this experiment, a polarized beam of neutrons interacted with the MTJ and MTJ-Ta. Subsequently, the spin and angle of the neutron beam reflected by the samples were analyzed.

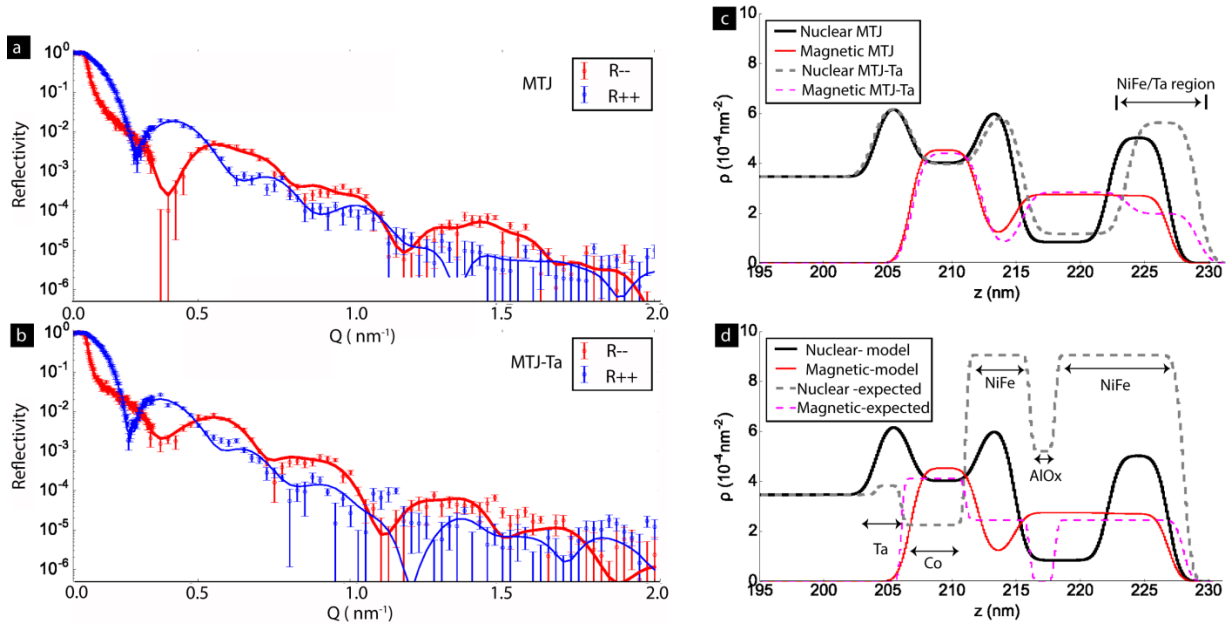

Fig. 5: Polarized beam neutron reflectivity study of (a) MTJ and (b) MTJ-Ta samples. Solid line show fitted curve on the experimental data. R-- correspond to reflectivity from nuclear profile minus reflectivity from magnetic profile. R++ corresponds to reflectivity from nuclear profile plus reflectivity from magnetic profile. (c) Nuclear and magnetic scattering length density (ρ) vs. MTJ and MTJ-Ta thickness. (d) Comparison of modeled and expected nuclear and magnetic scattering length density data for the MTJ.

We utilized nuclear reflectivity and non- spin flip reflectivity to calculate R-- and R++. Here, R-- corresponds to the difference in reflectivity from nuclear and magnetic profiles (Fig. 5). Whereas, the R++ correspond to the sum of reflectivity due to nuclear and magnetic profiles (Fig. 5). The R-- and R++ vs. wave vector graphs for the MTJ and MTJ-Ta are shown in Fig 5a and Fig 5b, respectively. The reflectivity profile for the MTJ (Fig. 5a) is different than that of MTJ-Ta (Fig. 5b). We fitted the experimental data with the scattering length density model to record the depth-wise changes in MTJ and MTJ-Ta magnetization (Fig. 5c). For MTJ-Ta magnetic signal in the 220 to 230 nm thickness range suggests that Ta gained magnetic moment. However, the magnetization in the adjacent NiFe's region decreased; the NiFe/Ta region for the magnetic signal in Fig. 5c indicates this possibility. However, for the MTJ sample magnetic signal for the top NiFe electrode has expected profile in the 220-230 nm thickness range (Fig. 5c). This PNR observation is in agreement with the prior studies which reported that a Ta film deposited on the NiFe gained magnetic moment [16, 17]. The reduction in magnetic moment of top NiFe electrode is also in agreement with the decrease in the intensity of the acoustic mode of MTJ as compared to MTJ-Ta (Fig. 2). Although Fig 5c represents the best fit for the nuclear and magnetic data, we do not believe the modeled data is perfectly accurate. To estimate the degree of deviation we compared the modeled data for the MTJ sample with the expected data (Fig. 5d). Expected data was calculated for the ideal MTJ with perfect interfaces and atomically smooth films. We attribute the difference between modeled and expected data to the significantly high roughness and diffusive interfaces between AlOx/NiFe (top) and NiFe(top)/Ta. Prior study [8] and our AFM study showed that growth of AlOx induced high roughness.

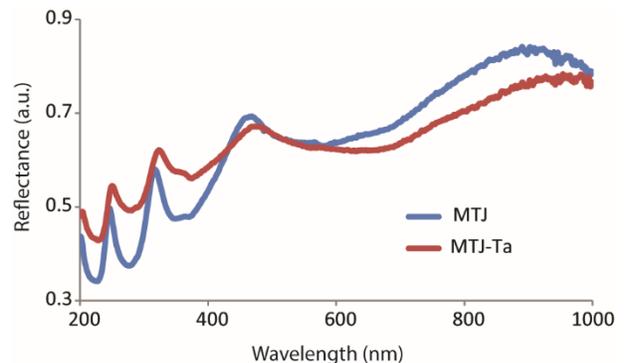

Fig. 6: Reflectance vs. wavelength graph for MTJ and MTJ-Ta.

In the quest of getting additional insights, we attempted to fit PNR data by fixing the thin film thickness in the scattering length density model (Fig.

1S, Supplementary material). The regions of nuclear and magnetic scattering length density for the bottom electrode (Ta/Co/NiFe) were in agreement with the expected thickness regions for the individual films. However, the nuclear and magnetic scattering length density did not show the good fit in sections corresponding to top NiFe and Ta films. This PNR study agrees with our hypothesis that issues mainly start after the deposition of AlOx. The PNR data provided in the Fig. 1S of the supplementary material indicated that magnetic moment was also present in the Ta region of MTJ-Ta. Future study may emphasize on producing smoother top AlOx and improved PNR modeling.

Even though roughness in the magnetic tunnel junctions impacted PNR modeling, but we do not believe this roughness level affected the integrity of magnetic tunnel junctions. The most delicate part of the magnetic tunnel junction is the AlOx tunneling barrier. We found that 3-5 nm level roughness is not detrimental to the integrity of tunnel barrier that separates the two ferromagnetic layers. For the validation, we produced tunnel junctions for the transport study by following the method described elsewhere [7]. The transport study conducted on tunnel junctions exhibited excellent tunneling behavior (Supplementary material-Fig.2S). The presence of tunneling response confirms that tunneling barrier is in good condition and unaffected by the level of roughness observed in our sample.

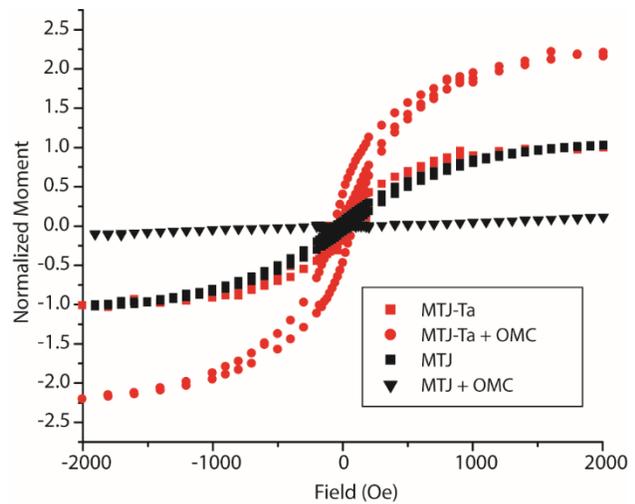

Fig. 7: Magnetic moment of MTJ and MTJ-Ta before and after treating with OMCs.

We conducted additional experiment to judge the quality of MTJ and MTJ-Ta. To make sure that top NiFe and Ta films were continuous we conducted light reflectivity study on MTJ and MTJ-Ta. We hypothesized that a continuous Ta film on the top of MTJ-Ta must produce clearly noticeable effects. The reflectivity data for MTJ and MTJ-Ta followed the similar trend (Fig. 7). The reflectivity data was consistent with the reflectivity profile from a continuous NiFe film [18]. However, the reflectivity data below ~450 nm was higher for the MTJ-Ta as compared to the MTJ sample (Fig. 6). It appears that Ta on top has a higher reflectivity for light radiation below ~450 nm only. We also noted that peaks of reflectivity data for MTJ-Ta were ~10 nm ahead of the MTJ samples for 200 to 450 nm range. Such a shift in the position of the reflectivity peaks for short wavelength is observed when the top NiFe film is covered with a conformal film of different composition [18]. This study suggests that although NiFe may be rough, but Ta has formed a conformal film on the top.

To demonstrate the application of maneuvering IFMEC with Ta, we studied the impact of Ta top layer on the magnetic properties of the magnetic tunnel junction based molecular devices (MTJMSD). The MTJMSD were produced by bridging the paramagnetic molecules between the ferromagnetic electrodes of the MTJ and MTJ-Ta [19]. These paramagnetic molecules are essentially organometallic molecular clusters (OMCs). For this study, the MTJ and MTJ-Ta samples discussed in Fig. 3, were utilized. The method of molecule attachment was described elsewhere [6, 7] . The attributes of OMC paramagnetic molecules have also been published elsewhere [20]. An OMC molecule contained an octa-nuclear cubic cage with a net spin state. Every corner of the OMC's cubic cage possessed an alkane tether. At the end of each alkane tether, a thiol functional group was provided. The thiol functional groups helped to bridge the molecules across the ~ 2 nm AlOx tunnel barrier along the exposed edges of the tunnel junctions. Each thiol group had a strong affinity towards the NiFe ferromagnetic electrode of the MTJ and MTJ-Ta. The magnetic study showed that OMCs created remarkably stronger exchange coupling as compared to AlOx tunnel barrier. It is noteworthy that MTJ and MTJ-Ta samples contained several thousand tunnel junctions to yield the high signal to noise ratio during magnetic studies. To study the paramagnetic molecule effect the magnetic moment were measured for ±2000 Oe field range at 150 K temperature. The magnetization study showed that OMC channels across the AlOx insulator on the exposed sides produced the opposite effects on MTJ and MTJ-Ta (Fig. 7). The magnetic moment of MTJ dropped nearly by ~84% (Fig. 7); it must be noted that MTJ had preexisting antiferromagnetic coupling between the ferromagnetic electrodes. On the other hand, OMCs increased the magnetic moment of the MTJ-Ta by ~116%; it is noteworthy that MTJ-Ta possessed pre-existing ferromagnetic coupling between the two ferromagnetic electrodes (Fig. 7). This study indicates that preexisting IFMEC is important in determining the magnetic properties of the molecular spintronics devices. We have conducted Monte Carlo study [19] to get qualitative understanding; however, simulation studies with continuous spin models are recommended to investigate the impact on preexisting IFMEC.

Here we propose the potential mechanism behind the Ta induced changes in the IFMEC of the MTJ. The previous observations of the reversal of IFMEC are reported due to variation in the property of the tunneling barriers[1]. However, it is noteworthy that in our study the tunneling barrier was grown with the same procedure for MTJ and MTJ-Ta. Hence, tunneling barrier is not expected to play a role in impacting IFMEC after the addition of Ta on MTJ. In our case, the magnetic interaction via the tunneling barrier is expected to be governed by the exchange coupling. Prior research showed that in the case of magnetic tunnel junctions exchange coupling was the most dominant [21] . The exchange coupling strength decreases exponentially[22] with the tunneling barrier thickness. IFMEC is also sensitive towards the crystallinity of the tunneling barrier[21, 23]. Our AlOx tunneling barrier growth method is based on prior work that leads to amorphous tunneling barrier[8]. In fact, AlOx tunneling barrier is by default amorphous[4, 23] and it is extremely challenging to get crystalline AlOx tunneling barrier [8].

Based on the experimental studies and prior literature we hypothesized the following mechanism behind the Ta effect on IFMEC. The reduced intensity of NiFe after the deposition of Ta may be due to the increased damping factor. We surmise that increase in damping may be associated with the creation of a dead layer at NiFe/Ta interface[16, 17, 24]. In this paper, we also proposed a mechanism based on the presences of a dead layer at NiFe/Ta interface [16, 17, 24]. Ta is found to acquire ~0.34-0.56 $\mu_B$ magnetic moment when deposited on NiFe surface [14,15]. These Ta atoms established antiferromagnetic coupling with the Ni atoms near the NiFe surface region. This antiferromagnetic coupling between the acquired magnetic moment in Ta and Ni atoms near the NiFe surface yielded a dead layer of ~ 2 nm thickness that does not possess any net magnetic moment [17]. Based on the prior study[14,15] the following mechanism is hypothesized about the Ta effect on IFMEC.

On a MTJ-Ta sample, the addition of Ta layer appears to gain magnetic moment by diminishing the spin density of NiFe (Fig. 8). Since, Ta layer acquires a net magnetic moment[14,15] hence it is imperative that Ta layer can only pick majority or minority spin density from NiFe, not both. If it picks both types of spins, then there may not be any net magnetic moment as reported by previous studies [14,15] and also seen in our PNR study (Fig. 5c).

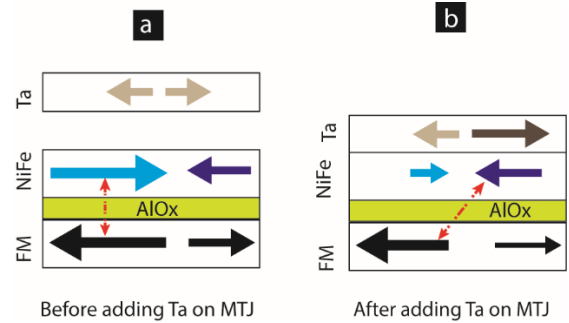

Fig.8: Ta effect on spin density of top ferromagnetic (FM) electrode and on the change in IFMEC.

Furthermore, we conjecture that FMR resonance peaks for the ferromagnets mainly depend on the majority spins only. In fact, FMR theoretical studies have mainly accounted for majority types of spin [25, 26] and ignored minority spin population. It is also well established that when two ferromagnets are coupled antiferromagnetically, then it means that the majority spins of the two ferromagnets are antiparallel to each other (Fig. 8a). For a MTJ sample, with antiferromagnetic IFMEC, the majority spin density (minority spin density) of the top NiFe electrode was antiparallel (parallel) to the majority spins of the bottom ferromagnetic electrode (Fig. 8a). We hypothesize that after addition of Ta, a fraction of the NiFe's majority spins moved into the Ta layer (Fig. 8b). As a result, NiFe's major spin density depleted and became lower than the minor spin density (Fig. 8b). Subsequently, the minor spin density of NiFe became new major spin density due to the presence of Ta (Fig. 8b). It is noteworthy that NiFe's new majority spin after Ta addition is parallel to the majority spins of the bottom electrode of the resultant MTJ-Ta. This new configuration is tantamount to ferromagnetic IFMEC on MTJ-Ta. Hence, the ferromagnetic coupling observed on MTJ-Ta is due to Ta induced rearrangement of the majority spin density of states on the top NiFe electrode (Fig. 8b). One can see that Ta effect is not possible if minority spins from the top NiFe enter in the Ta layer. In that case, NiFe's original majority spin remains unchanged before and after the addition of Ta layer. As a result, the nature of IFMEC will also not change. Conceptually, Ta induced IFMEC reversal is only possible when majority spins from the NiFe enter in the Ta layer. This hypothesis is in agreement with the difference in the intensities of the first peak for MTJ and MTJ-Ta. For MTJ-Ta the maximum intensity of the first peak is ~15191 around 598 Oe (Fig. 2). However, for MTJ the intensity of the first peak is 62720 at 620 Oe (Fig. 2). The addition of Ta appears to influence the population of spins responsible for producing the acoustic mode of the MTJ sample. According to FMR theory peak intensity is directly associated with the magnetic moment of the ferromagnetic electrodes[10, 14].

## IV CONCLUSIONS:

This paper discussed the effectiveness of Ta layer in changing the inter-ferromagnetic electrode exchange coupling. FMR was found to be useful in recording the subtle changes due to the addition of Ta layer. The magnetization study was only able to register very small change due to Ta. We also conducted neutron scattering studies on MTJ and MTJ-Ta samples. These neutron studies observed the moderate change in the magnetic attributes in the top Ta layer and neighboring NiFe region. We found that ferromagnetic resonance is extremely sensitive for studying the effect of the change in electrode composition on the inter-electrode exchange coupling. We found that the ability to change the nature of inter-electrode coupling of a magnetic tunnel junction can impact the resulting properties of the molecular spintronics devices. We observed that paramagnetic molecules decreased the magnetic moment of the MTJ with pre-existing antiferromagnetic coupling. However, the same paramagnetic molecules increased the magnetic moment of the MTJ-Ta with preexisting ferromagnetic exchange coupling. At present we are unsure about the mechanism by which the nature of Ta and NiFe interaction influences the IFMEC. First principle calculations are expected to shine light about the underlying mechanism.

**SUPPLEMENTARY MATERIAL**

Figure 1S showing the additional PNR results and Figure 2S showing the tunneling type transport via the AlOx tunnel barrier is provided in the supplementary material file.

**ACKNOWLEDGEMENTS:**

Pawan Tyagi thank Dr. Bruce Hinds and Department of Chemical and Materials engineering at the University of Kentucky for facilitating experimental work on molecular spintronics during his Ph.D. OMC was produced Dr. Stephen Holmes's group. The preparation of this paper and complementary experiments were in part supported by National Science Foundation-Research Initiation Award (Contract # HRD-1238802), Department of Energy/ National Nuclear Security Agency (Subaward No. 0007701-1000043016), and Air Force Office of Sponsored Research (Award #FA9550-13-1-0152). We also thank Centre of Nanoscience and Technology, NIST Gaithersburg for allowing the use of microscopy resources. We also acknowledge Dr. Brian Kirby of NIST Center of Neutron Reflectivity for the polarized beam reflectivity study. We also thank STEM Center at UDC for providing the partial funding. Any opinions, findings, and conclusions expressed in this material are those of the author(s) and do not necessarily reflect the views of any funding agency and corresponding author's affiliations and collaborators.